\documentclass[runningheads]{llncs}

\usepackage{graphicx}
\usepackage{adjustbox}
\usepackage{tikz}
\usepackage{xspace}
\usepackage{amssymb,amsmath}
\usepackage{times}
\usepackage[utf8]{inputenc}
\usepackage{array,booktabs}
\usepackage{longtable}
\usepackage{latexsym}
\usepackage{enumitem}
\usepackage{bbm}
\usepackage{multirow}
\usepackage{booktabs}
\usepackage{tabularx}
\usepackage{tikz}
\usetikzlibrary{arrows.meta, positioning, shapes, fit} 
\usepackage{pdflscape}
\pgfdeclarelayer{background}
\pgfsetlayers{background,main}
\usetikzlibrary{positioning}
\usepackage{seqsplit} 
\usepackage{changepage}

\usepackage{hyperref}
\hypersetup{
    colorlinks=false,
    linkcolor=blue,
    filecolor=magenta,      
    urlcolor=cyan,
    pdftitle={AI Act for the Working Programmer},
    pdfpagemode=FullScreen,
    }

\usepackage[misc]{ifsym}

\begin{document}

\title{Bridging the Disciplinary Gap in Explainable AI:\\ From Abstract Desiderata to Concrete Tasks}

\titlerunning{From Abstract Desiderata to Concrete Tasks}
\author{
Hanwei Zhang\inst{1}\and
Jingwen Wang\inst{2}\and
Holger Hermanns\inst{1}
}
\institute{
Saarland University, Saarland Informatics Campus, Saarbrücken, Germany\\
\email{{zhang,hermanns}@depend.uni-saarland.de}
\and
Universität zu Köln, Philosophische Fakultät, Ostasiatisches Seminar, Cologne, Germany\\
\email{jwang15@smail.uni-koeln.de}
}


\maketitle

\begin{abstract}
Explainable AI (XAI) is often criticized for failing to satisfy broad desiderata (e.g., fairness, accountability) and for limited practical value to stakeholders. This challenge partly arises because researchers across disciplines prioritize different sets of desiderata that remain underspecified and context-dependent, yet expect XAI to satisfy them simultaneously, resulting in fragmented and sometimes incompatible operationalizations.
We argue that many desiderata are not independent, but instead form dependency structures in which higher-level goals (\emph{e.g.}, trust, accountability) rely on more foundational properties (\emph{e.g.}, faithfulness, robustness). Some desiderata are multi-faceted and are best understood within these structures. In particular, instead of addressing all desiderata at once, we focus on subsets of dependency structures and translate them into concrete XAI tasks, thereby decomposing research questions into benchmarkable and solvable units.
To this end, we propose a three-axis taxonomy (\emph{target}, \emph{functional role}, and \emph{mode of justification}) and a three-step framework for deriving well-scoped, benchmarkable XAI tasks. Our approach builds on a systematic literature review and conceptual analysis, and supports clarifying desiderata, identifying dependencies, scoping feasibility, and delimiting the design space to derive concrete XAI tasks from abstract desiderata.
We illustrate its utility through two explanatory cases, showing how the taxonomy and framework guide systematic task design and evaluation in XAI. {\color{red}{This is a preprint of a paper that will appear in AISoLA 2026.}}
\end{abstract}

\section{Introduction}

Explainable AI (XAI) is increasingly critical as AI systems enter high-stakes domains like healthcare, finance, and criminal justice, promising to support trust, accountability, and regulatory compliance, including requirements under the EU AI Act~\cite{AIAct}. 
Despite numerous methods to generate explanations~\cite{zhang2021survey}, assess quality~\cite{zhang2024saliency}, and measure trust~\cite{hoffman2021measuring}, criticism has intensified that it fails to meet stakeholders’ desiderata, reflecting a persistent gap between technical developments and cross-disciplinary conceptualizations of these desiderata~\cite{ehsan2023charting}.
XAI faces a structural problem:
Large sets of research on desiderata have been proposed across disciplines, but existing approaches remain fragmented --- some focus on explanation fidelity \cite{doshi-velez2017towards}, others emphasize human interpretability without considering system or institutional constraints \cite{miller2019explanation}, and other efforts catalog desiderata without analyzing them or decomposing their underlying principles \cite{sokol2020explainable}.
Moreover, desiderata frequently remain ambiguous; although they often share similar labels, they refer to different explanatory properties, system-level qualities, or institutional outcomes, leading to conceptual overreach, misaligned evaluation, and difficulties in reasoning about trade-offs, assigning responsibility, and designing effective evaluation strategies.
Desiderata are often treated as independent, equally important properties that XAI is expected to satisfy simultaneously~\cite{fresz2024should}, making it difficult for researchers to identify a clear starting point for practical solutions.

Based on our experience and conceptual analysis, we argue that (1) many desiderata are multi-faceted and interdependent, and can be organized into dependency structures that specify which desiderata should be satisfied prior to others, with their meanings becoming more clear within such structures; and (2) desiderata should be decomposed into meaningful subsets, enabling large XAI research questions to be broken down into smaller, solvable units and translated into concrete tasks, thereby improving clarity, benchmarkability, and encouraging broader research engagement.

To this end, we propose a three-axis taxonomy and a three-step framework to help experts across disciplines derive well-defined XAI tasks aligned with the desiderata they value. Based on a literature review and conceptual analysis, the taxonomy distinguishes each desideratum along three dimensions: \emph{target}, \emph{functional role}, and \emph{mode of justification}. The target specifies what is constrained, such as the AI system, the explanation method, the human–system interface, or institutional governance. The functional role captures why it matters, distinguishing epistemic, pragmatic, protective, psychological, and normative functions. The mode of justification clarifies whether a desideratum is descriptive, instrumental, normative, or meta-normative. Together, these axes support structured reasoning about dependency, feasibility, and task design.

Building on this, we introduce a three-step framework: (1) map dependency relations among desiderata, (2) assess technical realizability, and (3) operationalize well-scoped XAI tasks. Modeling desiderata as nodes in directed dependency graphs distinguishes foundational from downstream desiderata and, together with feasibility analysis, helps identify where and why downstream desiderata fail to be satisfied. This, in turn, enables precise task formulation, supports the selection of evaluation metrics aligned with structurally necessary components rather than aspirational goals, and guides the decomposition of XAI research questions into solvable yet challenging units.
We illustrate the utility of the taxonomy and framework through two cases with which we are familiar and that we consider important: XAI for risk prevention in human oversight and XAI for accountability in auditing. These are not the only possible formulations and rely on our own analysis and judgment; however, we believe they are valuable both as explanatory cases and for applications of XAI in governance contexts.

\section{Background \& Related Work}
\vspace{-2ex} 
\paragraph{XAI Desiderata and Stakeholder-Oriented Taxonomies.}
A substantial body of XAI research explores desiderata that explanations or systems should satisfy, often grounded in stakeholder needs. Surveys such as those by Langer et al.~\cite{langer2021we} and Schwalbe and Finzel~\cite{schwalbe2024comprehensive} organize widely cited properties (\emph{e.g.}, faithfulness, fairness, robustness, trust, and accountability), highlighting how different actors prioritize distinct concerns. EUCA~\cite{jin2021euca} emphasizes end-user-centered desiderata, showing how user-relevant goals guide explanation design.
However, these taxonomies are largely descriptive: desiderata are presented as lists or clusters without modeling their interdependencies or the conditions under which they can be jointly satisfied. Consequently, desiderata at different levels, such as fidelity (technical), trust (psychological), and legal compliance (institutional), are often treated as coequal targets, obscuring that higher-level desiderata presuppose more foundational ones. Moreover, existing work lacks systematic tools for analyzing desiderata in a way that supports a clear, cross-disciplinary understanding.

\vspace{-2ex} 
\paragraph{Unifying Frameworks and Foundations for XAI.}
Prior work has proposed unified vocabularies and formal foundations for XAI~\cite{palacio2021xai,barbiero2023categorical,dembinsky2025unifying}, as well as human-centered and domain-specific perspectives~\cite{calzarossa2025assessment,broniatowski2021psychological,jin2021euca}.
Evaluation frameworks such as Doshi-Velez and Kim~\cite{doshi-velez2017towards} distinguish application-, human-, and functionally grounded evaluation, showing that explanation quality depends on context and user goals. Sokol and Flach~\cite{sokol2020explainable} propose structured assessment artifacts (“fact sheets”) to document assumptions and evaluation claims, overlapping with our aim of operationalizing and auditing tasks. 
While improving conceptual clarity, these approaches generally do not model dependencies among desiderata or support design-time reasoning about feasibility, trade-offs, and implementation.

Our work complements these approaches by providing a structured, operational methodology for analyzing how explanatory properties enable downstream outcomes, clarifying dependencies, feasibility, and task derivation.
We address these limitations by modeling desiderata as elements in dependency structures. We propose a three-axis taxonomy and a framework for deriving scoped XAI tasks. Unlike prior work, our approach supports design-time reasoning that links abstract desiderata to concrete tasks.

\providecommand*\emptycirc[1][0.5ex]{\tikz\draw (0,0) circle (#1);}
\providecommand*\halfcirc[1][0.5ex]{%
    \begin{tikzpicture}
    \draw[fill] (0,0)-- (90:#1) arc (90:270:#1) -- cycle ;
    \draw (0,0) circle (#1);
    \end{tikzpicture}}
\providecommand*\fullcirc[1][0.5ex]{\tikz\fill (0,0) circle (#1);}

\section{Taxonomy of Desiderata}
\label{sec:taxonomy}

We propose a taxonomy of desiderata to address the proliferation of overlapping, underspecified, and context-dependent concepts across disciplines. The taxonomy was derived through an iterative process combining a systematic literature review with conceptual analysis. We first identified XAI desiderata through a systematic search of relevant literature on Google Scholar using keywords such as “desiderata”, “explainable AI (XAI)”, “interpretability”, and “requirements”, consolidating study-specific labels into consistent conceptual entries (see Table~\ref{tab:vocabulary} for the full list). 
Through iterative abstraction, three dimensions emerged: \emph{target}, \emph{functional role}, and \emph{mode of justification}. This conceptual decomposition frames each desideratum as a constraint specifying what is constrained, for what purpose, and on what grounds. 

We map each desideratum onto these axes based on our interpretation of the source papers, assessing multifaceted cases by assigning relationship strengths (strong \fullcirc, intermediate \halfcirc, or weak \emptycirc) in Table~\ref{tab:vocabulary-complete}.
A \emph{strong} relation (\fullcirc) indicates that the desideratum is \emph{constitutively} defined by the axis category, reflecting a direct and consistently primary connection across the literature. An \emph{intermediate} relation (\halfcirc) indicates a meaningful but non-primary connection, where the desideratum contributes to the category but is more centrally characterized by another. A \emph{weak} relation (\emptycirc) captures incidental or context-dependent links that are not stable defining features. The absence of a symbol indicates no substantive connection.
For each desideratum, we examined its definitions, operationalizations, and empirical characterizations in the cited sources to assess its alignment with each axis. The classification was conducted collaboratively by authors from complementary disciplines (computer science and philosophy), with disagreements resolved through discussion and revisiting primary sources until consensus was reached. While the resulting judgments are necessarily interpretive, they provide a principled and informative reference for the field. In this section, we present the three-axis taxonomy and provide examples illustrating how classifications are derived from the literature.

\subsection{Axis 1: Target}

The first axis classifies XAI desiderata by the entity they aim to constrain, regulate, or improve, referred to as the \emph{target}. It makes explicit that satisfying a desideratum requires action of a specific entity, \emph{i.e.}, technical, human, or institutional, thereby clarifying who is responsible and which evaluation approaches are appropriate.

This axis addresses a common confusion: many desiderata are assessed via explanation methods but concern system-level properties, such as fairness, safety, or scientific understanding. Explanations serve as epistemic instruments, not the evaluation target. Conflating the two can lead to misdirected optimization, \emph{e.g.}, trying to fix system biases solely through explanation interfaces. Distinguishing what a desideratum is \emph{about} from how it is \emph{accessed} clarifies both technical feasibility and normative responsibility.
Operationally, targets are grouped into five categories: AI system, XAI method, explanation artifact, human interactions, and institutional or governance structures.

\textbf{System-centered (SYS)} desiderata (\emph{e.g.}, fairness, safety) constrain properties of the AI model itself; fulfilling them requires changes to the system's design or training. \textbf{XAI method-centered (XAI)} desiderata (\emph{e.g.}, faithfulness, scalability) evaluate whether an explanation technique reliably provides access to the system. \textbf{Explanation-centered (XAI-E)} desiderata (\emph{e.g.}, accessibility, complexity) evaluate the content of explanations as communicative objects. \textbf{Human-centered (HUM)} desiderata (\emph{e.g.}, trust, satisfaction) focus on the states of human agents. \textbf{Institutional-centered (INST)} desiderata (\emph{e.g.}, accountability, legal compliance) are realized through organizational procedures and social norms.

Together, these five targets provide a comprehensive identification
of the entities that XAI desiderata may constrain. The target axis generalizes across stakeholders and system components, enabling clear identification of loci of concern. For example, a deployer may consider desiderata at multiple levels: accountability and governance map to institution-centered targets, as they depend on external standards and regulations, whereas monitoring need XAI methods able to assess internal model behavior. By making explicit where intervention is required, the target axis avoids the common error of treating all XAI desiderata as properties of explanations alone.

\subsection{Axis 2: Functional Role}

Whereas Axis 1 identifies \emph{what entity or process a desideratum constrains}, Axis 2 classifies desiderata according to \emph{why they are required}, that is, the primary role they play in explainable AI. 
We refer to this dimension as the \emph{functional role}: the dominant contribution a desideratum makes to epistemic inquiry, practical decision-making, risk management, human experience, or normative justification.
Distinguishing functional roles is essential because XAI desiderata are often treated as pursuing a single objective, when they reflect fundamentally different aims and standards.
Even with a shared target, researchers may define success differently, assigning distinct functions to the same desideratum. For example, faithfulness may serve epistemic goals or foster user trust. Making functional roles explicit clarifies their purpose and evaluation criteria.

Operationally, the term \emph{primary function} follows functional analysis in philosophy of science and human-computer interaction, referring to the dominant, constitutive role a desideratum plays in XAI practice rather than all possible downstream effects~\cite{langer2021we,longo2024manifesto}. This does \emph{not} imply a desideratum has only one function; many are multifunctional and may serve multiple roles simultaneously, as reflected in Table~\ref{tab:vocabulary-complete} via strong (\fullcirc) or secondary (\halfcirc) associations across Axis~2 column. For instance, \emph{trust} serves both psychological (shaping user attitudes) and normative (legitimizing deployment decisions) functions. Thus, \emph{primary} distinguishes constitutive roles from incidental ones, clarifying what a desideratum is \emph{for} in context without reducing its functional richness.

The five following functional roles were derived from a cross-disciplinary synthesis of the XAI requirements literature~\cite{langer2021we,longo2024manifesto,navarro2021desiderata}, capturing recurring demand drivers across technical, human, and governance perspectives: 
\textbf{Epistemic (EPI)} functions concern producing and justifying knowledge about AI systems; \textbf{Action-guiding (ACT)} functions concern enabling intervention and decision-making; \textbf{Protective (PROT)} functions concern reducing harm and managing risks; \textbf{Psychological (PSY)} functions concern shaping human understanding and experience; and \textbf{Normative (NORM)} functions concern other supporting moral, legal, and institutional justification.
Support for this partition comes from interdisciplinary work on AI oversight.
Sterz et al.~\cite{sterz2024oversight}, for instance, derive the minimal conditions for effective human oversight as: \emph{epistemic access} (EPI), \emph{causal power} and \emph{self-control} (ACT), risk mitigation as the protective goal (PROT), individual factors such as automation bias, exhaustion, and motivation (PSY), and \emph{fitting intentions} together with the normative requirements of the EU AI Act (NORM). The convergence of 
developed analyses across computer science, psychology, law, and philosophy suggest that these five categories reflect a principled and non-arbitrary decomposition of XAI functional demands.

\subsection{Axis 3: Mode of Justification}

Axis 3 addresses \emph{how} a desideratum is justified: whether it is grounded in empirical facts, instrumental effectiveness, ethical values, or higher-order norms. This distinction is crucial to avoid category errors, such as trading off ethical obligations against technical metrics without normative reasoning.

\textbf{Descriptive/Technical (DES)} justification applies to desiderata grounded in measurable properties or formal proofs (\emph{e.g.}, accuracy, scalability). 
\textbf{Instrumental (INSTR)} justification applies to desiderata valued as means to an end (\emph{e.g.}, complexity, intuitiveness). 
\textbf{Normative (NORM)} justification applies to ethical or social commitments (\emph{e.g.}, fairness, autonomy). 
\textbf{Meta-normative (META)} justification regulates the legitimacy and scope of other requirements (\emph{e.g.}, legal compliance, informed consent). META justification manages the “normative pluralism” that arises from conflicting demands~\cite{baum2025alignment}, relying on a socio-technical ecosystem of legislation and jurisprudence~\cite{borges2022industry4} to ensure AI categories align with human dignity~\cite{sullivan2024hacking}.

\subsection{Illustrative Classification Examples from the Literature}
\label{subsec:examples}

To demonstrate the operational application of the classification criteria defined across the three axes, we present two representative examples drawn from the literature corpus. These cases illustrate how source studies are mapped onto relationship categories (Table~\ref{tab:vocabulary-complete}) through justifications grounded in our defined criteria.

\paragraph*{Example 1: Deck et al.~\cite{deck2024mapping} --- Fairness Auditing via XAI.}
Deck et al. analyze how XAI techniques relate to formal fairness definitions (individual, group, causal), constructing a cross-domain mapping between explanation properties and fairness criteria~\cite{deck2024mapping}.

\begin{itemize}
    \item \emph{Fairness} (\fullcirc\ SYS, PROT, META; \halfcirc\ ACT): The primary object is the AI system's compliance with fairness criteria (SYS) to prevent discriminatory harm (PROT). The meta-normative (META) assignment reflects the paper's central finding that different formal fairness definitions---such as predictive parity and equalized odds---are mutually incompatible and cannot be simultaneously satisfied~\cite{deck2024mapping}; selecting among them requires higher-order normative reasoning about which regulatory standard takes precedence in a given context, which is precisely the function of META justification. Fairness diagnostics guide model revision (ACT), which is considered a secondary relation.
    \item \emph{Monitoring}
    (\fullcirc\ XAI; \halfcirc\ NORM): XAI-based monitoring is explicitly invoked as the technical mechanism for bias detection and distributional drift across the AI lifecycle (XAI target). The secondary normative (NORM) assignment reflects the paper's framing of long-term monitoring as a requirement entailed by fairness obligations~\cite{deck2024mapping}: sustaining fairness over time is not merely a technical goal but a normatively mandated practice, making monitoring an action whose justification is ultimately normative rather than purely instrumental.
\end{itemize}

\paragraph*{Example 2: Sterz et al.~\cite{sterz2024oversight} --- Human Oversight and Accountability.}
Sterz et al. develop a framework for effective human oversight grounded in conditions for moral responsibility, requiring that the overseer possess epistemic access, retain causal agency, maintain self-control, and act with normatively suitable intentions~\cite{sterz2024oversight}. This example, to which we return in Section~\ref{sec:case_HO}, 
 informs several key assignments:

\begin{itemize}
    \item \emph{Responsibility} (\fullcirc\ SYS, EPI, INSTR): The paper argues that responsibility is realized at the system level through human--AI arrangements. Its core conditions are explicitly epistemic (sufficient knowledge) and instrumentally action-guiding (enabling intervention), satisfying the strong-relation criteria.
    \item \emph{Effectiveness} (\fullcirc\ HUM, ACT, INSTR; \halfcirc\ EPI, PROT): Effectiveness is conceptualized as the human capacity to correct failures through causal power and self-control---the oversight person must be able to intervene and follow through on risk-mitigating actions (ACT, HUM). Epistemic access and risk mitigation are discussed as necessary enabling conditions rather than primary definitional components of effectiveness itself.
\end{itemize}
These examples illustrate how our operational definitions produce evidence-anchored and replicable assignments. This same procedure was applied across the entries in the consolidated taxonomy.


\makeatletter
\newcommand{\tcite}[1]{%
  {\tiny\cite{#1}}%
}
\makeatother

\begin{adjustwidth}{-2cm}{-2cm}
\begingroup

\setlength{\LTleft}{0pt plus 1fill}
\setlength{\LTright}{0pt plus 1fill}

\scriptsize
\setlength{\tabcolsep}{1pt}
\begin{longtable}{@{\!\!\!\!\!\!\!\!\!\!\!\!\!\!\!\!\!\!\!\!\!\!\!\!\!\!\!\!\!\!\!\!\!\!\!\!\!\!\!\!\!\!\!\!}l|lllll|lllll|llll@{}}
\caption{\textbf{Vocabulary of Desiderata.} \fullcirc\ strong relation, \halfcirc\ secondary relation, \emptycirc\ weak/indirect relation, and no symbol indicates no substantive connection. Axis codes: SYS (system-centered), XAI (XAI method-centered), XAI-E (explanation-centered), HUM (human-centered), INST (institutional-centered), EPI (epistemic), ACT (action-guiding), PROT (protective), PSY (psychological), NORM (normative), DES (descriptive/technical), INSTR (instrumental), META (meta-normative). See Table~\ref{tab:vocabulary} for the desideratum definition.} 
\label{tab:vocabulary-complete}\\
\toprule
\multirow{2}{*}{\textbf{Desideratum}} 
& \multicolumn{5}{c|}{\textbf{Axis 1: Target}} 
& \multicolumn{5}{c|}{\textbf{Axis 2: Functional Role}}
& \multicolumn{4}{c}{\textbf{Axis 3: Mode of Justification}} \\
\cmidrule(lr){2-6} \cmidrule(lr){7-11} \cmidrule(lr){12-15}
&{SYS}&{XAI}&{XAI-E}&{HUM}&{INST}&{EPI}&{ACT}&{PROT}&{PSY}&{NORM}&{DES}&{INSTR}&{NORM}&{META}\\\midrule
\endfirsthead

\multicolumn{15}{c}{\tablename\ \thetable\ -- \textit{Continued from previous page}} \\
\toprule
\multirow{2}{*}{\textbf{Desideratum}} 
& \multicolumn{5}{c|}{\textbf{Axis 1: Target}} 
& \multicolumn{5}{c|}{\textbf{Axis 2: Functional Role}}
& \multicolumn{4}{c}{\textbf{Axis 3: Mode of Justification}} \\
\cmidrule(lr){2-6} \cmidrule(lr){7-11} \cmidrule(lr){12-15}
&{SYS}&{XAI}&{XAI-E}&{HUM}&{INST}&{EPI}&{ACT}&{PROT}&{PSY}&{NORM}&{DES}&{INSTR}&{NORM}&{META}\\\midrule
\endhead

\midrule
\multicolumn{15}{r}{\textit{Continued on next page}} \\
\endfoot

\bottomrule
\endlastfoot


Acceptance 
& \emptycirc\tcite{navarro2021desiderata} 
& 
& 
& \fullcirc\tcite{langer2021we,longo2024manifesto} 
& \halfcirc\tcite{aydin2024assessing,speith2025voodoobox,navarro2021desiderata} 
& 
& \emptycirc\tcite{speith2023smartproduction,langer2021we} 
& 
& \fullcirc\tcite{langer2021we,longo2024manifesto} 
& \halfcirc
& \emptycirc\tcite{langer2021we} 
& 
& \fullcirc\tcite{langer2021we,longo2024manifesto,navarro2021desiderata} 
& \\\midrule

Accountability
& \emptycirc\tcite{aryal2023even,heger2022understanding,navarro2021desiderata}
& 
& 
& \emptycirc
& \fullcirc\tcite{borges2022autonomous,langer2021we,longo2024manifesto}
& \emptycirc\tcite{navarro2021desiderata,langer2021we,longo2024manifesto}
& 
& 
& 
& \fullcirc\tcite{borges2022autonomous,langer2021we,longo2024manifesto}
& 
& 
& \fullcirc\tcite{borges2022autonomous,langer2021we,longo2024manifesto}
& \halfcirc
\\\midrule

Accuracy
& \fullcirc\tcite{hermanns2024aiact,borges2021productliability,langer2021we}
& \emptycirc
& 
& 
& 
& \fullcirc\tcite{hermanns2024aiact,borges2021productliability,langer2021we}
& \halfcirc
& 
& 
& 
& \fullcirc\tcite{langer2021we,kares2025saliencymap}
& 
& 
& 
\\\midrule

Actionability
& 
& 
& \fullcirc\tcite{bender2025towards,kastner2022psychopathology}
& \halfcirc
& 
& 
& \fullcirc\tcite{bender2025towards,kastner2022psychopathology}
& 
& 
& 
& 
& \fullcirc\tcite{bender2025towards,kastner2022psychopathology}
& 
& 
\\\midrule

Accessibility
& \emptycirc\tcite{alvarezromero2023governance}
& 
& \fullcirc\tcite{bruckner2025consent}
& \halfcirc
& \emptycirc\tcite{alvarezromero2023governance}
& 
& \fullcirc\tcite{alvarezromero2023governance,bruckner2025consent}
& 
& 
& 
& 
& \fullcirc\tcite{alvarezromero2023governance,bruckner2025consent}
& 
& 
\\\midrule

Autonomy
& 
& 
& 
& \fullcirc\tcite{mota2020altruism,aydin2024assessing,langer2021we}
& \emptycirc\tcite{walker2019planning,aydin2024assessing}
& 
& 
& 
& \fullcirc\tcite{mota2020altruism,aydin2024assessing,langer2021we}
& \halfcirc\tcite{walker2019planning,aydin2024assessing}
& 
& 
& \fullcirc\tcite{mota2020altruism,walker2019planning,aydin2024assessing}
& 
\\\midrule

Chronology
& 
& 
& \fullcirc\tcite{sokol2019desiderata}
& 
& 
& \fullcirc\tcite{sokol2019desiderata}
& \halfcirc
& 
& \emptycirc\tcite{sokol2019desiderata}
& 
& 
& 
& \fullcirc\tcite{sokol2019desiderata}
& 
\\\midrule

Complexity
& \halfcirc\tcite{mann2023opacity}
& 
& \fullcirc\tcite{sokol2019desiderata,kares2025saliencymap}
& \emptycirc\tcite{sokol2019desiderata,mann2023opacity}
& 
& \emptycirc\tcite{kares2025saliencymap,mann2023opacity}
& \fullcirc\tcite{sokol2019desiderata,kastner2022psychopathology}
& 
& 
& 
& \emptycirc\tcite{kares2025saliencymap}
& \fullcirc\tcite{sokol2019desiderata,kares2025saliencymap,mann2023opacity}
& 
& 
\\\midrule

Confidence
& \emptycirc\tcite{borges2021productliability}
& \emptycirc\tcite{speith2024conceptualizing}
& 
& \fullcirc\tcite{speith2024conceptualizing,longo2024manifesto}
& 
& 
& 
& \emptycirc\tcite{akoush2022serving}
& \fullcirc\tcite{speith2024conceptualizing,longo2024manifesto}
& \halfcirc
& \emptycirc\tcite{wollschlaeger2023uncertainty,akoush2022serving,ashmore2019assuring}
& 
& \fullcirc\tcite{speith2024conceptualizing,longo2024manifesto}
& 
\\\midrule

Contextualization 
& 
& 
& \fullcirc\tcite{sokol2019desiderata}
& 
& 
& \fullcirc\tcite{sokol2019desiderata}
& 
& 
& 
& 
& 
& \fullcirc\tcite{sokol2019desiderata}
& 
& 
\\\midrule

Controllability
& \halfcirc\tcite{acikgoz2025conversational}
& 
& 
& \fullcirc\tcite{acikgoz2025conversational,langer2021we,speith2025voodoobox}
& \emptycirc\tcite{borges2022autonomous}
& 
& \fullcirc\tcite{acikgoz2025conversational,langer2021we,speith2025voodoobox}
& \emptycirc\tcite{acikgoz2025conversational,borges2022autonomous,speith2025voodoobox}
& 
& 
& 
& \fullcirc\tcite{acikgoz2025conversational,langer2021we,speith2025voodoobox}
& \halfcirc\tcite{borges2022autonomous,speith2025voodoobox}
& 
\\\midrule

Debugability
& \emptycirc\tcite{speith2023xai}
& \fullcirc\tcite{speith2023xai}
& 
& 
& 
& 
& \fullcirc\tcite{speith2023xai}
& 
& 
& 
& 
& \fullcirc\tcite{speith2023xai}
& \halfcirc
\\\midrule

Discovery
& \halfcirc\tcite{kastner2024mechanistic}
& \fullcirc\tcite{langer2021we,kastner2024mechanistic}
& 
& \emptycirc\tcite{langer2021we}
& 
& \fullcirc\tcite{langer2021we,kastner2024mechanistic}
& \emptycirc\tcite{langer2021we}
& 
& 
& 
& 
& \fullcirc\tcite{langer2021we,kastner2024mechanistic}
& 
& 
\\\midrule

Distinction
& \fullcirc\tcite{baum2025taming,biewer2023doping}
& 
& 
& \emptycirc\tcite{baum2025taming,biewer2023doping}
& \halfcirc\tcite{baum2025taming,biewer2023doping}
& 
& 
& \fullcirc\tcite{baum2025taming,biewer2023doping}
& 
& \emptycirc\tcite{baum2025taming,biewer2023doping}
& 
& 
& \fullcirc\tcite{baum2025taming,biewer2023doping}
& 
\\\midrule

Education
& \emptycirc\tcite{pallocca2024tumorboards,heger2022understanding,baum2025taming}
& \emptycirc\tcite{borges2022autonomous,longo2024manifesto,sterz2025philosophers}
& \emptycirc\tcite{mann2023opacity}
& \fullcirc\tcite{pallocca2024tumorboards,heger2022understanding,biewer2023doping}
& \halfcirc\tcite{heger2022understanding,biewer2023doping,aydin2024assessing}
& \fullcirc\tcite{pallocca2024tumorboards,heger2022understanding,biewer2023doping}
& \emptycirc\tcite{pallocca2024tumorboards,heger2022understanding,biewer2023doping}
& \emptycirc\tcite{baum2025taming,borges2022autonomous,sullivan2024hacking}
& \halfcirc\tcite{biewer2023doping,aydin2024assessing,borges2022autonomous}
& \emptycirc\tcite{borges2022autonomous,sterz2025philosophers,speith2025voodoobox}
& \emptycirc\tcite{biewer2023doping,aydin2024assessing,hermanns2024aiact}
& \fullcirc\tcite{pallocca2024tumorboards,heger2022understanding,biewer2023doping}
& \emptycirc\tcite{borges2022autonomous,speith2025voodoobox}
& 
\\\midrule

Effectiveness
& \halfcirc\tcite{langer2021we,sterz2024oversight}
& \emptycirc\tcite{kares2025saliencymap,langer2021we,sterz2024oversight}
& \emptycirc\tcite{kares2025saliencymap}
& \fullcirc\tcite{kares2025saliencymap,langer2021we,sterz2024oversight}
& \emptycirc\tcite{langer2021we,sterz2024oversight}
& \halfcirc\tcite{kares2025saliencymap,langer2021we,sterz2024oversight}
& \fullcirc\tcite{langer2021we,sterz2024oversight}
& \halfcirc\tcite{langer2021we,sterz2024oversight}
& \emptycirc\tcite{langer2021we,sterz2024oversight}
& \emptycirc\tcite{langer2021we,sterz2024oversight}
& \emptycirc\tcite{kares2025saliencymap}
& \fullcirc\tcite{kares2025saliencymap,langer2021we,sterz2024oversight}
& \emptycirc\tcite{sterz2024oversight}
& \emptycirc\tcite{sterz2024oversight}
\\\midrule

Efficiency
& \fullcirc\tcite{langer2021we,longo2024manifesto,acikgoz2025conversational}
& \emptycirc\tcite{longo2024manifesto,navarro2021desiderata}
& \emptycirc\tcite{longo2024manifesto}
& \emptycirc\tcite{langer2021we,navarro2021desiderata}
& \halfcirc\tcite{lauber2025regulatory}
& \emptycirc\tcite{acikgoz2025conversational,navarro2021desiderata}
& \fullcirc\tcite{lauber2025regulatory,acikgoz2025conversational,navarro2021desiderata}
& \emptycirc\tcite{navarro2021desiderata,walker2019planning}
& \emptycirc\tcite{acikgoz2025conversational,walker2019planning}
& \emptycirc\tcite{lauber2025regulatory,navarro2021desiderata}
& \emptycirc\tcite{langer2021we,acikgoz2025conversational}
& \fullcirc\tcite{lauber2025regulatory,langer2021we,longo2024manifesto}
& \halfcirc\tcite{langer2021we,acikgoz2025conversational}
& \emptycirc\tcite{lauber2025regulatory,navarro2021desiderata}
\\\midrule

Fairness
& \fullcirc\tcite{biewer2023doping,deck2024mapping,langer2021we}
& \emptycirc\tcite{aydin2024assessing,baum2025taming,langer2021we}
& \emptycirc\tcite{aydin2024assessing,baum2025taming,langer2021we}
& \halfcirc\tcite{biewer2023doping,deck2024mapping,baum2025taming}
& \emptycirc\tcite{biewer2023doping,speith2024xhw,langer2021we}
& \halfcirc\tcite{aydin2024assessing,baum2025taming,langer2021we}
& \halfcirc\tcite{deck2024mapping,baum2025taming,langer2021we}
& \fullcirc\tcite{biewer2023doping,deck2024mapping,langer2021we}
& \emptycirc\tcite{biewer2023doping,speith2024xhw,langer2021we}
& \emptycirc\tcite{biewer2023doping,speith2024xhw,langer2021we}
& \emptycirc\tcite{biewer2023doping,langer2021we}
& \emptycirc\tcite{biewer2023doping,deck2024mapping,langer2021we}
& \fullcirc\tcite{biewer2023doping,speith2024xhw,langer2021we}
& \fullcirc\tcite{biewer2023doping,deck2024mapping,langer2021we}
\\\midrule

Faithfulness
& \emptycirc\tcite{kares2025saliencymap}
& \fullcirc\tcite{kares2025saliencymap,berlotattwell2021vqa}
& 
& 
& \emptycirc\tcite{kares2025saliencymap}
& \halfcirc
& 
& \emptycirc\tcite{kares2025saliencymap}
& 
& 
& 
& \fullcirc\tcite{kares2025saliencymap,berlotattwell2021vqa}
& 
& 
\\\midrule

Generalization
& \fullcirc\tcite{berlotattwell2021vqa,longo2024manifesto}
& \halfcirc\tcite{longo2024manifesto}
& 
& \emptycirc\tcite{langer2021we}
& 
& 
& 
& \emptycirc\tcite{longo2024manifesto}
& 
& 
& 
& \fullcirc\tcite{berlotattwell2021vqa,longo2024manifesto}
& 
& 
\\\midrule

Informed Consent
& \emptycirc\tcite{lauber2018research,speith2025voodoobox}
& 
& 
& \fullcirc\tcite{lauber2018research,speith2025voodoobox}
& \fullcirc\tcite{lauber2018research,speith2025voodoobox}
& 
& 
& \halfcirc\tcite{lauber2018research}
& 
& \halfcirc\tcite{lauber2018research,speith2025voodoobox}
& 
& 
& \fullcirc\tcite{lauber2018research,speith2025voodoobox}
& 
\\\midrule

Interpretability
& \fullcirc\tcite{ashmore2019assuring,speith2023smartproduction,freiesleben2023dear}
& \fullcirc\tcite{sterz2024oversight,speith2023smartproduction,freiesleben2023dear}
& \emptycirc\tcite{ashmore2019assuring,speith2023smartproduction}
& \halfcirc\tcite{sterz2024oversight,speith2023smartproduction,freiesleben2023dear}
& \emptycirc\tcite{ashmore2019assuring,freiesleben2023dear,kastner2024mechanistic}
& \fullcirc\tcite{sterz2024oversight}
& \emptycirc\tcite{ashmore2019assuring}
& 
& 
& \emptycirc\tcite{ashmore2019assuring,freiesleben2023dear,kastner2024mechanistic}
& 
& \fullcirc\tcite{ashmore2019assuring,sterz2024oversight,speith2023smartproduction}
& \emptycirc\tcite{freiesleben2023dear,kastner2024mechanistic}
& 
\\\midrule

Legal Compliance
& \emptycirc\tcite{langer2021we,speith2024xhw}
& \emptycirc\tcite{langer2021we}
& \emptycirc\tcite{langer2021we}
& \emptycirc\tcite{langer2021we}
& \fullcirc\tcite{langer2021we,speith2024xhw}
& 
& 
& \fullcirc\tcite{langer2021we,speith2024xhw}
& 
& \halfcirc\tcite{langer2021we,speith2024xhw}
& \emptycirc\tcite{langer2021we}
& 
& \fullcirc\tcite{langer2021we,speith2024xhw}
& 
\\\midrule

Monitoring
& \emptycirc\tcite{biewer2023doping,bruckner2025consent,baum2025taming}
& \fullcirc\tcite{baum2025taming,deck2024mapping}
& \emptycirc\tcite{baum2025taming,deck2024mapping}
& \emptycirc\tcite{biewer2023doping,bruckner2025consent,baum2025taming}
& \emptycirc\tcite{lauber2025regulatory,hermanns2024aiact,baum2025taming}
& \emptycirc\tcite{biewer2023doping,baum2025taming}
& \fullcirc\tcite{biewer2023doping,hermanns2024aiact,baum2025taming}
& \halfcirc\tcite{biewer2023doping,bruckner2025consent,hermanns2024aiact}
& \emptycirc\tcite{biewer2023doping,baum2025taming}
& \emptycirc\tcite{biewer2023doping,baum2025taming}
& \fullcirc\tcite{lauber2025regulatory,hermanns2024aiact,baum2025taming}
& \fullcirc\tcite{biewer2023doping,bruckner2025consent,hermanns2024aiact}
& \halfcirc\tcite{biewer2023doping,hermanns2024aiact,deck2024mapping}
& \emptycirc\tcite{baum2025taming,deck2024mapping}
\\\midrule

Morality/Ethics
& \emptycirc\tcite{langer2021we,baum2025alignment}
& \emptycirc\tcite{langer2021we}
& \emptycirc\tcite{langer2021we}
& \emptycirc\tcite{langer2021we,baum2025alignment}
& \fullcirc\tcite{langer2021we,baum2025alignment}
& 
& 
& \fullcirc\tcite{langer2021we,baum2025alignment}
& 
& \halfcirc\tcite{langer2021we,baum2025alignment}
& \emptycirc\tcite{langer2021we}
& 
& \fullcirc\tcite{langer2021we,baum2025alignment}
& 
\\\midrule

Performance
& \fullcirc\tcite{bender2025towards,langer2021we,longo2024manifesto}
& \halfcirc\tcite{bender2025towards,langer2021we,longo2024manifesto}
& \emptycirc\tcite{kares2025saliencymap}
& \emptycirc\tcite{baum2025taming,kares2025saliencymap}
& \emptycirc\tcite{baum2025taming,speith2023smartproduction,lauber2025regulatory}
& \emptycirc\tcite{baum2025taming,kares2025saliencymap}
& \fullcirc\tcite{bender2025towards,langer2021we,longo2024manifesto}
& \emptycirc\tcite{baum2025taming,speith2023smartproduction}
& 
& 
& \emptycirc\tcite{baum2025taming,lauber2025regulatory,borges2021productliability}
& \halfcirc\tcite{bender2025towards,kares2025saliencymap}
& \fullcirc\tcite{bender2025towards,baum2025taming,langer2021we}
& \emptycirc\tcite{lauber2025regulatory,borges2021productliability}
\\\midrule

Privacy
& \fullcirc\tcite{aydin2024assessing,lauber2018research,deck2024mapping}
& \halfcirc\tcite{aydin2024assessing}
& \emptycirc\tcite{aydin2024assessing,lauber2018research}
& \emptycirc\tcite{lauber2018research}
& \emptycirc\tcite{aydin2024assessing,lauber2018research,deck2024mapping}
& \fullcirc\tcite{lauber2018research}
& \emptycirc\tcite{aydin2024assessing,lauber2018research}
& 
& 
& \halfcirc\tcite{aydin2024assessing,lauber2018research,deck2024mapping}
& 
& \fullcirc\tcite{aydin2024assessing,lauber2018research}
& \emptycirc\tcite{aydin2024assessing,lauber2018research,deck2024mapping}
& 
\\\midrule

Responsibility
& \fullcirc\tcite{sterz2024oversight}
& \halfcirc\tcite{speith2023smartproduction,sullivan2024hacking}
& \emptycirc\tcite{sterz2024oversight,speith2023smartproduction,sullivan2024hacking}
& \emptycirc\tcite{sterz2024oversight,speith2023smartproduction}
& \emptycirc\tcite{sterz2024oversight,speith2023smartproduction}
& \fullcirc\tcite{sterz2024oversight}
& \emptycirc\tcite{sterz2024oversight,speith2023smartproduction}
& 
& 
& \halfcirc\tcite{sterz2024oversight,speith2023smartproduction,sullivan2024hacking}
& 
& \fullcirc\tcite{sterz2024oversight,speith2023smartproduction,sullivan2024hacking}
& \emptycirc\tcite{sterz2024oversight,speith2023smartproduction,sullivan2024hacking}
& 
\\\midrule

Robustness (AI)
& \fullcirc\tcite{biewer2023doping,hermanns2024aiact,baum2025alignment}
& \emptycirc\tcite{hermanns2024aiact}
& \emptycirc\tcite{aydin2024assessing,hermanns2024aiact}
& \emptycirc\tcite{hermanns2024aiact}
& \emptycirc\tcite{hermanns2024aiact,baum2025alignment}
& \fullcirc\tcite{biewer2023doping,aydin2024assessing,hermanns2024aiact}
& \halfcirc\tcite{biewer2023doping,baum2025alignment}
& 
& 
& \emptycirc\tcite{aydin2024assessing,hermanns2024aiact,baum2025alignment}
& 
& \fullcirc\tcite{biewer2023doping,aydin2024assessing,hermanns2024aiact}
& \emptycirc\tcite{hermanns2024aiact,baum2025alignment}
& 
\\\midrule

Robustness (XAI)
& \emptycirc\tcite{kares2025saliencymap,longo2024manifesto}
& \fullcirc\tcite{kares2025saliencymap,sullivan2024hacking,longo2024manifesto}
& \halfcirc\tcite{kares2025saliencymap,sullivan2024hacking,longo2024manifesto}
& \emptycirc\tcite{kares2025saliencymap,longo2024manifesto}
& \emptycirc\tcite{kares2025saliencymap,longo2024manifesto}
& \fullcirc\tcite{kares2025saliencymap,sullivan2024hacking,longo2024manifesto}
& \emptycirc\tcite{kares2025saliencymap,longo2024manifesto}
& 
& 
& \emptycirc\tcite{sullivan2024hacking,longo2024manifesto}
& 
& \fullcirc\tcite{kares2025saliencymap,longo2024manifesto}
& \emptycirc\tcite{sullivan2024hacking,longo2024manifesto}
& 
\\\midrule

Safety
& \fullcirc\tcite{hermanns2024aiact,borges2021productliability,borges2022industry4}
& \emptycirc\tcite{longo2024manifesto}
& \emptycirc\tcite{longo2024manifesto}
& \emptycirc\tcite{borges2021productliability,longo2024manifesto}
& \emptycirc\tcite{hermanns2024aiact,borges2021productliability,borges2022autonomous}
& \fullcirc\tcite{borges2021productliability,baum2025alignment,longo2024manifesto}
& \halfcirc\tcite{hermanns2024aiact,borges2021productliability,borges2022industry4}
& 
& 
& \emptycirc\tcite{hermanns2024aiact,borges2021productliability,borges2022autonomous}
& 
& \fullcirc\tcite{hermanns2024aiact,borges2021productliability,borges2022industry4}
& \emptycirc\tcite{hermanns2024aiact,borges2021productliability,borges2022autonomous}
& 
\\\midrule

Satisfaction
& \emptycirc\tcite{langer2021we}
& \emptycirc\tcite{langer2021we}
& \emptycirc\tcite{kares2025saliencymap}
& \fullcirc\tcite{kares2025saliencymap}
& 
& \halfcirc\tcite{langer2021we,speith2024xhw}
& 
& 
& \fullcirc\tcite{kares2025saliencymap}
& 
& \emptycirc\tcite{langer2021we}
& \emptycirc\tcite{kares2025saliencymap}
& 
& \fullcirc\tcite{langer2021we,kares2025saliencymap,speith2024xhw}
\\\midrule

Scalability
& \halfcirc\tcite{biewer2023doping,bruckner2025consent,baum2025taming}
& 
& \fullcirc\tcite{nemec2025xai}
& \emptycirc\tcite{biewer2023doping,baum2025taming}
& \emptycirc\tcite{bruckner2025consent}
& 
& \fullcirc\tcite{biewer2023doping,bruckner2025consent,baum2025taming}
& 
& 
& \emptycirc\tcite{bruckner2025consent}
& 
& \fullcirc\tcite{biewer2023doping,bruckner2025consent,baum2025taming}
& \emptycirc\tcite{bruckner2025consent}
& 
\\\midrule

Science
& \fullcirc\tcite{lauber2018research,kastner2024mechanistic}
& \fullcirc\tcite{longo2024manifesto,kastner2024mechanistic,mann2023opacity}
& \emptycirc\tcite{longo2024manifesto,mann2023opacity}
& \halfcirc\tcite{lauber2018research,kastner2024mechanistic,mann2023opacity}
& \emptycirc\tcite{lauber2018research,longo2024manifesto,deck2024mapping}
& \fullcirc\tcite{longo2024manifesto,kastner2024mechanistic,mann2023opacity}
& \emptycirc\tcite{lauber2018research,kastner2024mechanistic}
& 
& 
& \emptycirc\tcite{lauber2018research,longo2024manifesto,deck2024mapping}
& 
& \fullcirc\tcite{lauber2018research,longo2024manifesto,kastner2024mechanistic}
& \emptycirc\tcite{lauber2018research,deck2024mapping,kastner2024mechanistic}
& 
\\\midrule

Security
& \fullcirc\tcite{baum2025alignment,speith2025voodoobox,speith2024xhw}
& 
& 
& \emptycirc\tcite{speith2025voodoobox,speith2024xhw}
& \emptycirc\tcite{borges2022industry4}
& 
& 
& \fullcirc\tcite{borges2022industry4,baum2025alignment,speith2025voodoobox}
& \emptycirc\tcite{speith2025voodoobox,speith2024xhw}
& \halfcirc\tcite{borges2022industry4,speith2025voodoobox,speith2024xhw}
& \emptycirc\tcite{speith2025voodoobox,speith2024xhw}
& 
& \fullcirc\tcite{borges2022industry4,baum2025alignment}
& 
\\\midrule

Sufficiency
& \emptycirc\tcite{borges2022autonomous}
& 
& \fullcirc\tcite{speith2024conceptualizing,borges2022autonomous}
& \halfcirc\tcite{speith2024conceptualizing,borges2022autonomous}
& \emptycirc\tcite{borges2022autonomous}
& \fullcirc\tcite{speith2024conceptualizing,borges2022autonomous}
& \emptycirc\tcite{speith2024conceptualizing}
& 
& \emptycirc\tcite{speith2024conceptualizing}
& \emptycirc\tcite{speith2024conceptualizing,borges2022autonomous}
& \emptycirc\tcite{borges2022autonomous}
& \emptycirc\tcite{borges2022autonomous}
& \fullcirc\tcite{speith2024conceptualizing,borges2022autonomous}
& \halfcirc\tcite{speith2024conceptualizing}
\\\midrule

Transparency
& \emptycirc\tcite{speith2024xhw,mann2023opacity}
& \emptycirc\tcite{deck2024mapping,mann2023opacity}
& \emptycirc\tcite{langer2021we,mann2023opacity}
& \emptycirc\tcite{aydin2024assessing,langer2021we}
& \fullcirc\tcite{aydin2024assessing,bruckner2025consent,hermanns2024aiact}
& \fullcirc\tcite{aydin2024assessing,langer2021we,deck2024mapping}
& \emptycirc\tcite{aydin2024assessing,bruckner2025consent,hermanns2024aiact}
& 
& 
& \halfcirc\tcite{aydin2024assessing,bruckner2025consent,hermanns2024aiact}
& \emptycirc\tcite{speith2024xhw,mann2023opacity}
& 
& \fullcirc\tcite{aydin2024assessing,bruckner2025consent,hermanns2024aiact}
& \emptycirc\tcite{deck2024mapping,mann2023opacity}
\\\midrule

Trust
& \emptycirc\tcite{baum2025taming,langer2021we,kastner2021trust}
& \fullcirc\tcite{baum2025taming,borges2022industry4,langer2021we}
& \emptycirc\tcite{borges2022industry4,kastner2021trust}
& \halfcirc\tcite{baum2025taming,kares2025saliencymap,langer2021we}
& \emptycirc\tcite{kares2025saliencymap,langer2021we,speith2023smartproduction}
& \emptycirc\tcite{baum2025taming,borges2022industry4}
& \fullcirc\tcite{baum2025taming,borges2022industry4,langer2021we}
& \emptycirc\tcite{baum2025taming,borges2022industry4,kastner2021trust}
& \halfcirc\tcite{kares2025saliencymap,langer2021we,speith2023smartproduction}
& 
& \emptycirc\tcite{baum2025taming,borges2022industry4,kastner2021trust}
& 
& 
& 
\\\midrule

Trustworthiness
& \fullcirc\tcite{baum2025taming,langer2021we,kastner2021trust}
& \emptycirc\tcite{baum2025taming,langer2021we,kastner2021trust}
& \emptycirc\tcite{baum2025taming,kastner2021trust}
& \fullcirc\tcite{baum2025taming,langer2021we,kastner2021trust}
& \emptycirc\tcite{speith2023smartproduction}
& \emptycirc\tcite{baum2025taming,kastner2021trust}
& 
& \fullcirc\tcite{baum2025taming,kastner2021trust}
& \halfcirc\tcite{baum2025taming,langer2021we}
& 
& \fullcirc\tcite{baum2025taming,kastner2021trust}
& 
& 
& 
\\\midrule

Understandability
& \fullcirc\tcite{sterz2024oversight,aydin2024assessing,bender2025towards}
& 
& 
& \halfcirc\tcite{sterz2024oversight,aydin2024assessing,bender2025towards}
& \emptycirc\tcite{aydin2024assessing}
& \fullcirc\tcite{sterz2024oversight,aydin2024assessing,bender2025towards}
& \emptycirc\tcite{aydin2024assessing,bender2025towards}
& \emptycirc\tcite{sterz2024oversight,aydin2024assessing}
& \emptycirc\tcite{sterz2024oversight,bender2025towards}
& \halfcirc\tcite{aydin2024assessing,bender2025towards}
& \fullcirc\tcite{sterz2024oversight,bender2025towards}
& 
& \fullcirc\tcite{aydin2024assessing,bender2025towards}
& 
\\\midrule

Usability
& \fullcirc\tcite{bruckner2025consent,langer2021we,speith2025voodoobox}
& \emptycirc\tcite{bruckner2025consent,langer2021we,kares2025saliencymap}
& \emptycirc\tcite{bruckner2025consent}
& \halfcirc\tcite{langer2021we}
& \emptycirc\tcite{langer2021we,speith2025voodoobox}
& \emptycirc\tcite{bruckner2025consent,speith2025voodoobox}
& \emptycirc\tcite{bruckner2025consent,kares2025saliencymap,speith2025voodoobox}
& \emptycirc\tcite{bruckner2025consent,langer2021we}
& \fullcirc\tcite{bruckner2025consent,langer2021we,kares2025saliencymap}
& 
& \emptycirc\tcite{bruckner2025consent,langer2021we}
& 
& 
& 
\\\midrule

Verification
& \fullcirc\tcite{bruckner2025consent,hermanns2024aiact,langer2021we}
& \fullcirc\tcite{chitgopkar2024accuracy}
& 
& \emptycirc\tcite{bruckner2025consent,langer2021we,speith2023smartproduction}
& \halfcirc\tcite{hermanns2024aiact,speith2024xhw}
& \fullcirc\tcite{bruckner2025consent,hermanns2024aiact,speith2024xhw}
& \emptycirc\tcite{bruckner2025consent,langer2021we,speith2024xhw}
& \emptycirc\tcite{speith2023smartproduction,langer2021we}
& \emptycirc\tcite{speith2023smartproduction,hermanns2024aiact,speith2024xhw}
& \halfcirc\tcite{bruckner2025consent,hermanns2024aiact,langer2021we}
& \fullcirc\tcite{bruckner2025consent,langer2021we,speith2023smartproduction}
& 
& \fullcirc\tcite{hermanns2024aiact,langer2021we,speith2023smartproduction}
& 
\\\midrule

Verifiability
& \fullcirc\tcite{speith2024xhw,longo2024manifesto}
& 
& \fullcirc\tcite{nemec2025xai}
& \emptycirc\tcite{speith2024xhw}
& \halfcirc\tcite{speith2024xhw,longo2024manifesto}
& \fullcirc\tcite{speith2024xhw,longo2024manifesto}
& \emptycirc\tcite{speith2024xhw}
& \halfcirc\tcite{speith2024xhw,longo2024manifesto}
& \emptycirc\tcite{longo2024manifesto}
& \emptycirc\tcite{speith2024xhw,longo2024manifesto}
& \fullcirc\tcite{speith2024xhw}
& 
& \emptycirc\tcite{speith2024xhw,longo2024manifesto}
& 
\\\bottomrule

\end{longtable}
\endgroup
\end{adjustwidth}

\normalsize


\section{From Desiderata to Scoped XAI Tasks: A Taxonomy-Guided Guideline}

Satisfying all XAI desiderata simultaneously is neither technically feasible nor conceptually possible. Advancing XAI, therefore, requires decomposing desiderata into well-defined tasks that can be clearly specified and communicated across disciplines. In the previous section, we introduced a taxonomy to characterize desiderata by their target, functional role, and mode of justification. Building on this, we now present a guideline to support experts in using the taxonomy to derive concrete XAI tasks, thereby enabling researchers to operationalize desiderata in ways that are tailored to specific contexts and interests. This guideline serves as a conceptual bridge between underspecified stakeholder desiderata and narrowly defined technical tasks, facilitating cross-disciplinary communication and enabling the systematic translation of abstract goals into implementable XAI solutions.

As desiderata are defined in relation to stakeholder needs, it is essential to first identify the stakeholder’s purpose, often corresponding to downstream desiderata. Given that desiderata exhibit dependency relations, the initial step is to map the dependencies surrounding these downstream desiderata. We recommend using the existing vocabulary of desiderata as shown in Table~\ref{tab:vocabulary}. The next step is to assess technical feasibility across the identified dependencies to locate key bottlenecks. Finally, these challenges should be translated into concrete XAI tasks, specified as precisely as possible to guide implementation. To clarify the proposed guideline and its application, we first outline the procedure, followed by two illustrative cases that are familiar to the authors and of particular importance and urgency in light of the EU AI Act. 
The two dependency structures in Fig.~\ref{fig:huamnoversight-relationships} and Fig.~\ref{fig:xai-framework} we shall derive for the illustrative cases are only one of many possible analyses and are \emph{not} unique.

\subsection{Guideline Overview: A Three-Step Task-Derivation Framework}

We propose a three-step framework for translating abstract XAI desiderata into concrete, well-scoped tasks. First, we identify the specific downstream desiderata and map their dependencies on related desiderata. Second, we assess technical feasibility to locate key challenges in achieving these downstream desiderata. Third, we specify tasks that are both feasible and normatively grounded. Together, these steps provide a principled workflow for designing, evaluating, and situating XAI methods within broader socio-technical contexts.

\vspace{-1.7ex}
\paragraph{Step 1: Identifying Dependency Relations Among Desiderata.}

We begin by identifying a small set of key downstream desiderata, ideally grounded in a concrete scenario. Expert judgment is essential for scoping and managing the task. From these downstream goals, relevant desiderata can be identified using the existing vocabulary, with the functional role axis providing the most informative starting point. Dependencies among desiderata should then be established, where the target and mode of justification axes are particularly informative. For example, XAI-related (XAI, XAI-E) desiderata generally presuppose system-centered (SYS) desiderata, which themselves are grounded in human and institutional (HUM, INST) desiderata, while descriptive and instrumental (DES, INSTR) desiderata often ground normative and meta-normative (NORM, META) ones.
These dependencies can be represented as directed edges, where a source desideratum must be satisfied for a target desideratum to be meaningfully pursued. Such edges capture epistemic, technical, and interactional prerequisites rather than strict logical or causal relations. Operationally, they indicate the minimal conditions required for downstream goals to be achievable. Experts should iteratively construct, reflect on, and refine this dependency structure through discussion until a stable consensus is reached.

\vspace{-1.7ex}
\paragraph{Step 2: Assessing Technical Feasibility.}

The second step assesses which segments of each dependency chain are technically realizable. Here, “technical” extends beyond XAI algorithms to include methods appropriate to each target. For instance, system-centered desiderata (SYS) require grounding in AI models, human-centered targets (HUM) involve human–computer interaction and cognitive considerations, and normative or meta-normative justifications (NORM, META) draw on philosophical analysis, while institutional targets (INST) may involve legal or governance frameworks. The three axes indicate relevant domains for assessing technical feasibility. 

Once the relevant techniques are identified, feasibility is evaluated against the literature along four criteria: \emph{epistemic accessibility, formalizability, algorithmic implementability, and evaluability}. \emph{Epistemic} accessibility refers to whether a desideratum has been sufficiently analyzed and articulated with a clear definition. \emph{Formalizability} concerns whether it has been expressed in formal terms, such as mathematical formulations. \emph{Algorithmic implementability} assesses whether existing computational methods can realize the desideratum. \emph{Evaluability} considers whether there are clear metrics to determine whether, and to what extent, the desideratum is satisfied. Experts can define additional feasibility levels as appropriate for their tasks along these four criteria. Low feasibility does not imply a poor XAI method or an invalid desideratum. A desideratum may still be satisfied in a targeted and adequate way even if it is not fully operationalized across all four criteria.

We distinguish three feasibility levels; while not exhaustive, they are simple and useful for the two illustrative cases: (1) \emph{operationalizable}, where all criteria are satisfied, and the desiderata can be directly implemented and evaluated given sufficient concrete information; (2) \emph{mechanism-dependent}, where feasibility requires introducing new mechanisms (e.g., in the AI system, XAI methods, or institutional processes); and (3) \emph{downstream}, where higher-level desiderata may be unattainable when their prerequisite mechanisms are not yet feasible; they cannot currently be achieved by XAI alone, though XAI may still be a crucial component of broader solutions.

\vspace{-1.7ex}
\paragraph{Step 3: Deriving XAI Tasks.}
The final step translates the dependency and feasibility analyses into concrete XAI tasks. This involves identifying (1) what is missing to fulfill 
XAI to fulfill the downstream desiderata, and (2) the disciplines best suited to address them. In practice, mechanism-dependent desiderata identified in the previous step are the primary focus, and the target axis helps indicate the relevant domains and expertise required. XAI tasks should be specified as precisely as possible and refined through cross-disciplinary collaboration to ensure clarity, feasibility, and alignment with the underlying desiderata.


\subsection{Explanatory Cases I: XAI Task for Human Oversight to Prevent Risk}
\label{sec:case_HO}

To illustrate the practical use of our taxonomy and three-step framework, we first consider the case of supporting deployer-oriented desiderata for human oversight in risk prevention. We select this case due to its urgency in regulatory contexts such as the EU AI Act~\cite{AIAct}, as well as its practical importance. We believe that enabling XAI in this setting is both feasible and impactful.

\begin{figure}[tpbh]
\centering
\scriptsize
\makebox[\linewidth][c]{
\begin{tikzpicture}[
    node distance=0.4cm and 1.0cm,
    every node/.style={
        draw,
        rectangle,
        rounded corners=2pt,
        minimum width=2.2cm,
        minimum height=0.7cm,
        align=center,
        font=\small,
        fill=white,
        inner sep=3pt,
        line width=0.6pt
    },
    arrow/.style={
        -Stealth,
        line width=0.6pt,
        shorten >=2pt,
        shorten <=2pt,
        color=black!70
    }
]

\node (discrimination) at (0,0) {Distinction};
\node (chronology) [below=of discrimination] {Chronology};
\node (effectiveness) [below=of chronology] {Effectiveness};
\node (faithfulness) [below=of effectiveness, fill=teal!15] {Faithfulness};
\node (contextfulness) [right=of discrimination] {Contextualization};
\node (understanding) [right=of contextfulness] {Understandability};
\node (complexity) [right=of understanding, fill=teal!15] {Complexity};
\node (actionability) [below=of contextfulness, fill=teal!15] {Actionability};
\node (accessibility) [below=of actionability, fill=teal!15] {Accessibility};
\node (monitory) [below=of accessibility, fill=teal!15] {Monitoring};
\node (controllability) [right=of monitory, fill=teal!15] {Controllability};
\node (robustness-xai) [below=of faithfulness, fill=teal!15] {Robustness\\(XAI)};
\node (efficiency) [right=of accessibility, fill=teal!15] {Efficiency \\(Timeless)};
\node (responsibility) [right=of controllability, fill=orange!25] {Responsibility};
\node (robustness-ai) [right=of robustness-xai, fill=teal!15] {Robustness\\(AI)};
\node (security-safety) [right=of robustness-ai] {Resilience \\ (Security \& Safety)};
\node (performance) [right=of security-safety, fill=orange!25] {Performance};

\draw[arrow] (effectiveness.east) -- (monitory.west);
\draw[arrow] (monitory) -- (controllability);
\draw[arrow] (controllability) -- (responsibility);
\draw[arrow] (effectiveness.east) -- (contextfulness.west);
\draw[arrow] (contextfulness) -- (actionability);
\draw[arrow] (actionability.east) -- (controllability.west);
\draw[arrow] (accessibility) -- (actionability);
\draw[arrow] (efficiency.north) -- (actionability.east);
\draw[arrow] (discrimination) -- (contextfulness);
\draw[arrow] (contextfulness) -- (understanding);
\draw[arrow] (understanding.south) -- (responsibility.north);
\draw[arrow] (complexity) -- (understanding);
\draw[arrow] (chronology.east) -- (contextfulness.west);
\draw[arrow] (faithfulness.east) -- (robustness-ai.west);
\draw[arrow] (robustness-ai) -- (security-safety);
\draw[arrow] (security-safety) -- (performance);
\draw[arrow] (robustness-xai) -- (robustness-ai);

\node[draw, fill=orange!25, minimum width=0.4cm, minimum height=0.3cm, inner sep=0pt, rounded corners=1pt, font=\scriptsize] (lg1) at (-0.6,-5.8) {};
\node[draw=none, anchor=west, font=\scriptsize] at (-0.3,-5.8) {Downstream};

\node[draw, fill=teal!15, minimum width=0.4cm, minimum height=0.3cm, inner sep=0pt, rounded corners=1pt, font=\scriptsize] (lg2) at (3.3,-5.8) {};
\node[draw=none, anchor=west, font=\scriptsize] at (3.6,-5.8) {Operationalizable};

\node[draw, fill=white, minimum width=0.4cm, minimum height=0.3cm, inner sep=0pt, rounded corners=1pt, font=\scriptsize] (lg3) at (6.6,-5.8) {};
\node[draw=none, anchor=west, font=\scriptsize] at (7.1,-5.8) {Mechanism-dependent};

\end{tikzpicture}
}
\normalsize
\caption{\textbf{Dependency structure of desiderata for human oversight in risk prevention.} Nodes represent desiderata. \emph{Orange} nodes indicate downstream desiderata; \emph{teal} nodes indicate operationalizable desiderata supported by existing literature; \emph{white} nodes indicate mechanism-dependent desiderata requiring additional mechanisms, scope specification, and metrics for feasibility. Arrows point from more foundational to more dependent desiderata.}
\label{fig:huamnoversight-relationships}
\end{figure}
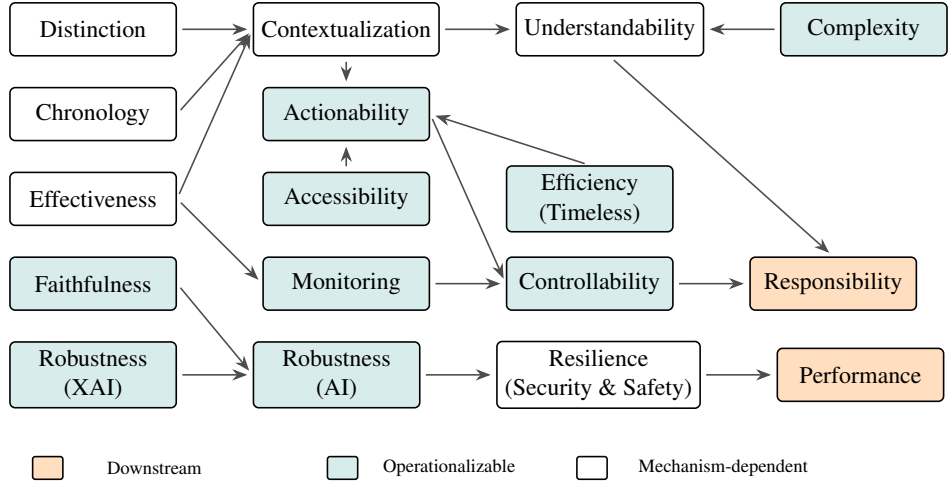


\vspace{-1.7ex}
\paragraph{Step 1.}
We first identify the minimal downstream desiderata for human oversight based on our understanding: deployers require XAI to enable humans to detect and reject potentially ill-intentioned inputs and to correct system errors. This leads to two key desiderata, namely \emph{responsibility} and \emph{performance}.
On the human-oversight side, \emph{responsibility} requires that the overseer satisfy several conditions~\cite{sterz2024oversight}: sufficient epistemic access to the decision context, self-control over available courses of action, causal influence over outcomes, and intentions aligned with their institutional role.
Based on our review of the desiderata vocabulary and analysis of their functional roles, we derive a set of XAI desiderata that enable these conditions and relate to the two key downstream desiderata, as summarized in Fig.~\ref{fig:huamnoversight-relationships}.
For instance, in human oversight scenarios, explanations must support context-aware interpretation, timely action, and effective human control. We translate these requirements into the XAI desiderata of \emph{understandability, contextualization, actionability, efficiency, effectiveness} and \emph{controllability}, capturing the need for interpretable, intervention-supporting explanations that sustain human oversight.

Then we identify dependencies by recognizing that these desiderata presuppose more foundational ones. We first analyze their targets and position XAI and XAI-E at a more foundational level, followed by system-centered (SYS) and then human-centered (HUM) desiderata. Through discussion and reflection, we find that XAI methods must be \emph{accessible} and \emph{efficient} in operational settings, enable effective \emph{monitoring} of system states, and provide \emph{effective}, temporally consistent (\emph{chronological}) information to support reliable causal inference. These properties ground higher-level oversight capabilities in implementable mechanisms. We further observe that improvements in \emph{performance} depend on \emph{security} and \emph{safety}, indicating that performance gains are contingent on these protective conditions. In turn, \emph{resilience (security $\&$ safety)} relies on the \emph{robustness} of the AI system, which is ultimately grounded in \emph{XAI robustness}.

\vspace{-1.7ex}
\paragraph{Step 2.}
Given the detailed structure, we then determine the technical feasibility of each desiderata. We find that most desiderata target SYS, XAI, and XAI-E, which are closely aligned with computer science. Based on a review of the literature, we identify that
\emph{faithfulness} can be evaluated through perturbation or counterfactual consistency tests~\cite{doshi-velez2017towards,zhang2024opti,petsiuk2018rise}; \emph{robustness (XAI)} and \emph{robustness (AI)} can be stress-tested under distributional shift~\cite{hendrycks2019benchmarking}; \emph{complexity} can be achieved by sparse constraint~\cite{cranmer2021disentangled}; \emph{monitoring} can be operationalized via anomaly detection or uncertainty estimation~\cite{lakshminarayanan2017simple,hendrycks2017baseline}; and \emph{actionability} can be assessed via predefined intervention protocols~\cite{christiano2017deep,shin2023closer}. \emph{Distinction} and \emph{chronology} have been sufficiently analyzed, with existing mathematical formulations and evaluation methods; however, their algorithmic implementation within XAI remains underdeveloped. For example, \emph{chronology} requires incorporating causal structure, which is largely absent in current XAI approaches.
\emph{Effectiveness, contextualization}, and \emph{understandability} are widely discussed but lack precise definitions, leaving their formalization, algorithmic implementation, and evaluation unclear.
\emph{Resiliency (i.e., security $\&$ safety)} is well-defined and implemented at the system level without XAI, but its role as an XAI desideratum remains largely unexplored.

\vspace{-1.7ex}
\paragraph{Step 3.}
To translate the dependency structure into XAI tasks, we first need to identify the key difficulties. In general, failures to satisfy downstream desiderata arise from unresolved mechanism dependencies, and each mechanism-dependent desideratum requires corresponding technical effort. In this illustrative case, an additional challenge emerges: although many desiderata are individually operationalizable, their underlying techniques may conflict. For instance, most desiderata are implemented by relying on attribution-based XAI, whereas \emph{actionability} is more naturally supported by concept-based methods~\cite{koh2020concept,zhang2026sl}. Aligning the techniques underlying different desiderata is a non-trivial task.
Thus, we identify the following key difficulties: (1) XAI methods and potentially AI systems require mechanisms for causal reasoning to support chronology; (2) XAI methods require mechanisms to support distinction; (3) coordination is needed across techniques associated with different operationalizable desiderata; (4) psychological work is needed to clarify what constitutes effective and context-sensitive explanations; (5) human–computer interface mechanisms are required to ensure understandability and effectiveness of existing techniques; and (6) interdisciplinary analysis is needed to clarify how XAI can improve the resilience (security $\&$ safety) of AI systems.

We translate these challenges into a concrete XAI task, including the following sub-tasks:
\begin{itemize}
    \item \textbf{Causal XAI for temporal reasoning}: Develop XAI methods that incorporate causal structure to generate explanations that preserve temporal and causal dependencies.
    \item \textbf{Discriminative explanation mechanisms}: Design XAI methods that enable a clear distinction between relevant and irrelevant factors, improving the selectivity and precision of explanations.
    \item \textbf{Multi-desiderata alignment}: Develop frameworks or pipelines that coordinate and reconcile potentially conflicting XAI techniques (e.g., attribution-based vs. concept-based methods) to jointly satisfy multiple desiderata.
    \item \textbf{Operationalizing explanation effectiveness and context}: Define and validate measurable criteria for Effectiveness and Contextualization through empirical and psychological studies, enabling their formalization and evaluation.
    \item \textbf{Interface-aware XAI design}: Design human–computer interaction mechanisms that ensure explanations are understandable and actionable in real-world settings, integrating XAI outputs into user-centered interfaces.
    \item \textbf{XAI for system resilience}: Investigate how XAI can contribute to improving the security and safety of AI systems, including identifying vulnerabilities, supporting monitoring, and enhancing robustness.
\end{itemize}

\subsection{Explanatory Cases: XAI Task for Legal Audit}

As the first illustrative case is primarily oriented toward computer science techniques, \emph{i.e.,} focusing on SYS, XAI, and XAI-E targets, we provide a second example that is equally urgent in regulatory contexts and practically important, but emphasizes human (HUM) and institutional (INST) targets. This case concerns XAI desiderata for auditing. Unlike human oversight, which requires real-time intervention, auditing operates across temporal, organizational, and domain boundaries.



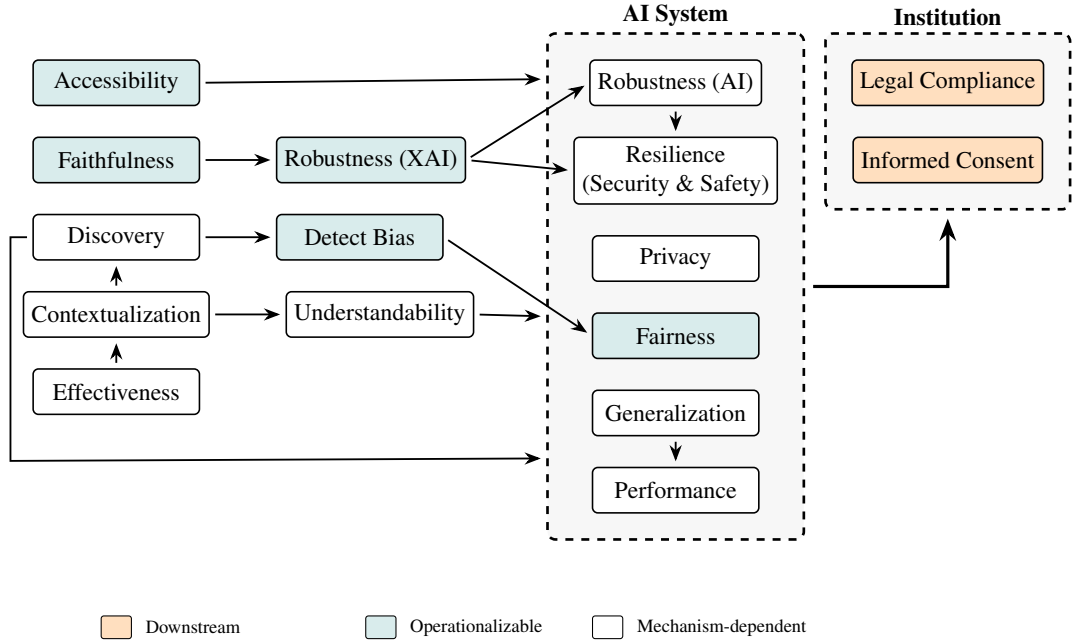
\begin{figure}[htbp]
\centering
\scriptsize
\makebox[\linewidth][c]{
\begin{tikzpicture}[
    node distance=0.4cm and 1.0cm,
    concept/.style={
        draw,
        rectangle,
        rounded corners=2pt,
        minimum width=2.2cm,
        minimum height=0.6cm,
        align=center,
        font=\small,
        fill=white,
        inner sep=3pt,
        line width=0.6pt
    },
    group/.style={
        draw,
        rectangle,
        rounded corners=3pt,
        line width=1pt,
        inner sep=10pt,
        fill=black!3,
        dashed
    },
    arrow/.style={
        -Stealth,
        line width=0.7pt,
        shorten >=2pt,
        shorten <=2pt
    },
    group arrow/.style={
        -Stealth,
        line width=1.2pt,
        shorten >=3pt,
        shorten <=3pt
    },
    label/.style={
        font=\small\bfseries
    }
]
\begin{scope}[local bounding box=xai]
\node[concept, fill=teal!15] (accessibility) {Accessibility};
\node[concept, below=of accessibility, fill=teal!15] (faithfulness) {Faithfulness};
\node[concept, below=of faithfulness] (discovery) {Discovery};
\node[concept, right=of faithfulness, fill=teal!15] (robxai) {Robustness (XAI)};
\node[concept, below=of discovery] (contextfullness) {Contextualization};
\node[concept, right=of contextfullness] (understanding) {Understandability};
\node[concept, below=of contextfullness] (effectiveness) {Effectiveness};
\node[concept, right=of discovery, fill=teal!15] (detectbias) {Detect Bias};
\end{scope}

\begin{scope}[shift={(7.4,0)}, local bounding box=ai]
\node[concept] (robai) {Robustness (AI)};
\node[concept, below=of robai] (security) {Resilience \\ (Security \& Safety)};
\node[concept, below=of security] (privacy) {Privacy};
\node[concept, below=of privacy, fill=teal!15] (fairness) {Fairness};
\node[concept, below=of fairness] (generalization) {Generalization};
\node[concept, below=of generalization] (performance) {Performance};
\end{scope}

\begin{scope}[shift={(11,0)}, local bounding box=inst]
\node[concept, fill=orange!25] (legal) {Legal Compliance};
\node[concept, below=of legal, fill=orange!25] (informed) {Informed Consent};
\end{scope}

\begin{pgfonlayer}{background}
\node[group, fit=(ai), label={[label, anchor=north]north:AI}] (box2) {};
\node[group, fit=(inst), label={[label, anchor=north]north:Institution}] (box3) {};
\end{pgfonlayer}

\node[label, anchor=south] at (box2.north) {AI System};
\node[label, anchor=south] at (box3.north) {Institution};

\draw[arrow] (faithfulness) -- (robxai);
\draw[arrow] (effectiveness) -- (contextfullness);
\draw[arrow] (discovery) -- (detectbias);
\draw[arrow] (contextfullness) -- (understanding);
\draw[arrow] (contextfullness) -- (discovery);

\draw[arrow] (robai) -- (security);
\draw[arrow] (generalization) -- (performance);

\draw[arrow] (robxai.east) -- (robai.west);
\draw[arrow] (robxai.east) -- (security.west);
\draw[arrow] (detectbias.east) -- (fairness.west);

\draw[arrow] (accessibility.east) -- (box2.122);

\draw[arrow] (discovery.west) -- ++(-0.2,0) -| (-1.4,-5) -- (box2.233);
\draw[arrow] (understanding.east) -- (box2.193);
\draw[group arrow] (box2.east) -| (box3.south);

\node[draw, fill=orange!25, minimum width=0.4cm, minimum height=0.3cm, inner sep=0pt, rounded corners=1pt] (lg1) at (0,-7.2) {};
\node[draw=none, anchor=west, font=\scriptsize] at (0.3,-7.2) {Downstream};

\node[draw, fill=teal!15, minimum width=0.4cm, minimum height=0.3cm, inner sep=0pt, rounded corners=1pt] (lg2) at (3.5,-7.2) {};
\node[draw=none, anchor=west, font=\scriptsize] at (3.8,-7.2) {Operationalizable};

\node[draw, fill=white, minimum width=0.4cm, minimum height=0.3cm, inner sep=0pt, rounded corners=1pt] (lg3) at (6.5,-7.2) {};
\node[draw=none, anchor=west, font=\scriptsize] at (6.8,-7.2) {Mechanism-dependent};

\end{tikzpicture}
}
\normalsize
\caption{\textbf{Dependency structure of desiderata for Legal Audit.} Nodes represent desiderata. \emph{Orange} nodes indicate downstream desiderata; \emph{teal} nodes indicate operationalizable desiderata supported by existing literature; \emph{white} nodes indicate mechanism-dependent desiderata requiring additional mechanisms, scope specification, and metrics for feasibility. Arrows point from foundational to dependent desiderata. Dashed boxes indicate groups; inter-group arrows mean full group-to-group dependency, while arrows to a group indicate the source is a prerequisite for all desiderata in that group.}
\label{fig:xai-framework}
\end{figure}

\vspace{-1.7ex}
\paragraph{Step 1.}

In our scenario, we exclude voluntary certification and focus on mandatory audits required for market approval. In this context, we identify the downstream desiderata as \emph{legal compliance} and \emph{informed consent}. We then reviewed the desiderata list alongside regulatory frameworks such as the EU AI Act~\cite{AIAct} to identify requirements imposed on AI systems, and incorporated the relevant desiderata into the dependency structure shown in Fig.~\ref{fig:xai-framework}.
When analyzing dependencies among desiderata, we examine them along the target axis. We find that the relations among SYS, XAI, and XAI-E are relatively clear. Between SYS and INST, we observe that SYS-level desiderata must be satisfied prior to INST-level ones. For clarity of representation, we group SYS desiderata into a single dashed box and INST desiderata into another, and depict their fully connected one-way dependency using a single directed arrow in Fig.~\ref{fig:xai-framework}.

We argue that \emph{accessibility} and the ability of XAI to support knowledge acquisition about the AI system (\emph{i.e., discovery}), as well as \emph{understandability} for humans, are prerequisites for all SYS-level desiderata. This follows from the specific scenario: the auditing organization relies on XAI to obtain sufficient information to justify whether the AI system satisfies required conditions.
Although SYS desiderata, such as \emph{resilience}, appear in both dependency structures, their functional roles differ. In the human oversight case, \emph{resilience}
is classified as protective (PROT), whereas in the auditing context, it is interpreted as epistemic (EPI) within the dependency structure.

\vspace{-1.7ex}
\paragraph{Step 2.}

We assess the feasibility of these desiderata along four criteria: epistemic accessibility, formalizability, algorithmic implementability, and evaluability. Within the XAI domain, properties such as \emph{faithfulness}, \emph{accessibility}, \emph{robustness (XAI)}, and \emph{detect bias} are both formalizable and evaluable. In the AI domain, most desiderata already have established metrics independent of XAI; however, how XAI can substantively demonstrate or certify these properties remains largely unexplored, with \emph{fairness} being a notable exception.

Beyond computational aspects, mechanism-dependent desiderata also require inputs from legal, philosophical, and psychological domains. Substantial epistemic work is still needed to clarify both individual desiderata and their interrelations. For instance, how institution-centered (INST) desiderata grounded in meta-normative justification can guide system-centered (SYS) desiderata framed in normative terms, or how explanations can provide sufficient information about AI system properties to support auditing.



\vspace{-1.7ex}
\paragraph{Step 3.}

Focusing on the “blank” regions of feasibility, we attempt to identify the most urgent challenges for advancing XAI toward informative explanations that support diagnostic metrics and system-level evaluation in the form of structured audit reports documenting verified links, assumptions, limitations, and evidential support for each claim.
This requires addressing four key questions: (1) regulatory stakeholders, in collaboration with researchers across disciplines, must determine the minimum requirements for satisfying normative constraints and assess to what extent AI systems meet them; (2) interdisciplinary work is needed to clarify how XAI can capture or reflect properties such as robustness, resilience, privacy, generalization, and performance, similar to how \cite{deck2024mapping} relates XAI techniques to formal notions of fairness; (3) technical mechanisms must be developed to enable XAI systems to discover and verify knowledge about AI system properties; and (4) cognitive evaluation is required to define what constitutes contextual, effective, and understandable explanations in auditing settings.

Thus, we translate them into an XAI task, including the following sub-tasks:
\begin{itemize}
    \item \textbf{Normative requirement operationalization for auditing}: Develop frameworks that enable regulatory stakeholders, in collaboration with researchers, to formalize minimum normative requirements and assess the extent to which AI systems satisfy them.
    \item \textbf{XAI–property mapping for system-level attributes}: Establish principled mappings between XAI techniques and system-level properties such as robustness, resilience, privacy, generalization, and performance, extending prior work such as \cite{deck2024mapping} on fairness.
    \item \textbf{Mechanism design for XAI-based system verification}: Develop technical mechanisms that enable XAI systems to discover, extract, and verify knowledge about internal AI system properties in a reliable and auditable manner.
    \item \textbf{Cognitive grounding of audit-oriented explanations}: Define and empirically validate criteria for what constitutes contextual, effective, and understandable explanations in auditing environments, grounded in human cognitive and decision-making constraints.
\end{itemize}

\subsection{Discussion}

The three-axis taxonomy supports the full procedure from abstract desiderata to concrete XAI tasks. In Step 1, it enables systematic construction of dependency structures by separating target, clarifying purpose (functional role), and distinguishing normative status (mode of justification), making presuppositional relations explicit. In Step 2, it guides feasibility analysis by linking targets to relevant function roles or modes of justification and clarifying required mechanisms. In Step 3, it helps identify gaps in dependency chains and translate them into concrete, solvable tasks. 

Across both explanatory cases, dependency structures are derived by combining literature evidence with interpretive analysis of presupposition relations. They should therefore be understood as structured conceptual models rather than empirically learned dependencies.
We aim to propose XAI tasks that inspire further research on the two scenarios of interest and support the development of XAI for stakeholders in these contexts. However, they reflect the authors’ interpretations and experience. While we believe they are valuable to the field, we also encourage others to propose alternative formulations and further refine these tasks.

The framework is inherently interpretive: both desiderata classification and dependency construction rely on expert judgment, so the taxonomy does not yield a unique or empirically fixed structure and may admit multiple valid decompositions. It also lacks large-scale empirical validation. Future work should empirically test dependency structures via expert elicitation, annotation studies, or domain-specific benchmarks to improve robustness and reproducibility. Despite these limitations, the framework provides a structured basis for standardizing XAI desiderata, making dependencies explicit and supporting more principled benchmarking, clearer cross-disciplinary communication, and systematic task design aligned with technical and stakeholder requirements.
\section{Conclusion}

This paper addresses a structural problem in XAI: cross-disciplinary desiderata are often treated as equally important and independent without a clear specification of their dependencies, justifications, targets, and functions. This leads to conceptual ambiguity, misaligned evaluation, and unclear feasibility, making it difficult to identify research entry points.
To address this, we propose a three-axis taxonomy and a three-step framework that enable precise specification of desiderata, analysis of dependency structures, feasibility assessment, and derivation of well-scoped XAI tasks.
We demonstrate the framework in two scenarios: human oversight and legal audit. While dependent on authors’ interpretation and experience, the examples show how the approach clarifies desiderata, exposes feasibility constraints, and supports task formulation, potentially inspiring further work in these areas.
Future work will focus on empirically validating and extending this framework, and we invite the community to contribute. Key directions include testing dependency structures via expert elicitation and annotation studies, developing benchmarks from dependency-based tasks, and building automated tools for mapping desiderata into the three-axis taxonomy. We also encourage applications in domains such as healthcare and finance, quantification of feasibility notions, and studies of how stakeholders interpret and use dependency structures.

\section*{Acknowledgements}
This work received support from DFG under grant No.~389792660 as part of TRR~248, as well as by DFG grant No.~547583482. Besides, we sincerely thank Sarah Sterz for taking the time to provide constructive feedback and valuable suggestions.

\bibliographystyle{splncs04}
\bibliography{reference}


\appendix
\scriptsize
\begin{longtable}{p{2.8cm}p{2.5cm}p{6.2cm}c}
\toprule
\textbf{Desideratum} & \textbf{Other Name(s)} & \textbf{Description} & \textbf{Cite} \\
\midrule
\endfirsthead

\multicolumn{4}{c}%
{\tablename\ \thetable\ -- \textit{Continued from previous page}} \\
\toprule
\textbf{Desideratum} & \textbf{Other Name(s)} & \textbf{Description} & \textbf{Cite} \\
\midrule
\endhead

\midrule
\multicolumn{4}{r}{\textit{Continued on next page}} \\
\endfoot

\endlastfoot

Acceptance & Adoption, Willingness-to-use & Users are willing to adopt, accept, and continuously use a system in practice & \cite{langer2021we} \\
\addlinespace

Accountability & Traceability & Provide appropriate evidence—such as data, models, decision processes, permissions, and logs—that enables determining who is responsible for system outcomes and under which rules responsibility is assigned & \cite{langer2021we} \\
\addlinespace

Accuracy & -- & Assess and increase a system's predictive accuracy; minimize errors & \cite{langer2021we,hunsicker2025100216} \\
\addlinespace

Actionability & - & An explanation provides implementable intervention cues that enable users to modify inputs, workflows, thresholds, or model configurations to change outcomes or reduce risk & \cite{nemec2025xai,navarro2021desiderata,sokol2019desiderata} \\
\addlinespace

Accessibility & Availability & An explanation must be easily, prominently, and adequately available & \cite{colmenarejo2025should} \\
\addlinespace

Autonomy & -- & Enable humans to retain their autonomy when interacting with a system & \cite{langer2021we} \\
\addlinespace

Chronology & -- & More recent causes of an event are preferred & \cite{sokol2019desiderata} \\
\addlinespace

Complexity & Sparseness, Parsimony & Explanation should minimize structural complexity and cognitive burden while preserving essential information & \cite{colmenarejo2025should,sokol2019desiderata,bender2025towards} \\
\addlinespace

Confidence & -- & Make humans confident when using a system; users feel confident when interacting with a system, particularly their subjective belief that they can correctly use the system and make appropriate decisions based on its outputs & \cite{langer2021we} \\
\addlinespace

Contextualization & Contextual adequacy, Contextfulness & An explanation satisfies completeness and provides the necessary conditions under which the foil holds contextfulness & \cite{navarro2021desiderata,sokol2019desiderata} \\
\addlinespace

Controllability & -- & Retain (complete) human control concerning a system & \cite{langer2021we} \\
\addlinespace

Debugability & -- & The extent to which the system and its explanations support identifying error sources (data, features, code, model components, or workflows) and verifying the effectiveness of fixes & \cite{speith2023xai} \\
\addlinespace

Detecting bias & -- & XAI should help to identify biases in data, representations, or decision rules & \cite{deck2024mapping,rong2023towards} \\
\addlinespace

Discovery & Exploring, Gain Knowledge & An XAI method should allow data scientists to discover patterns or connections that have not been utilized by other approaches; these patterns or connections can offer new insights in domain knowledge & \cite{renftle2022explaining} \\
\addlinespace

Distinction & Discrimination, Disentanglement & XAI should present the reasoning behind the result, which can help identify the source for the discrimination and address it accordingly & \cite{wang2024desiderata} \\
\addlinespace

Education & Learning support & Learn how to use a system and the system's peculiarities & \cite{langer2021we} \\
\addlinespace

Effectiveness & -- & Assess and increase a system's effectiveness; work effectively with a system & \cite{langer2021we,deters2024explainable} \\
\addlinespace

Efficiency & -- & The extent to which task objectives are achieved with minimal resource consumption & \cite{langer2021we,deters2024explainable} \\
\addlinespace

Fairness & -- & Grounded in the general normative framework of non-discrimination and fair treatment & \cite{langer2021we,deck2024mapping} \\
\addlinespace

Faithfulness & Fidelity, Soundness, Correctness & The extent to which the provided explanatory information reflects the system's actual decision-making process & \cite{speith2023xai,colmenarejo2025should,bender2025towards,nemec2025xai,deters2024explainable} \\
\addlinespace

Generalization & -- & An XAI method should explain novel data points & \cite{lipton2018mythos} \\
\addlinespace

Homogeneity & -- & An XAI method maintains comparable reliability across demographic groups (gender, age, and race/ethnicity) & \cite{colmenarejo2025should} \\
\addlinespace

Informed Consent & -- & Enable humans to give their informed consent concerning a system's decisions & \cite{langer2021we} \\
\addlinespace

Interpretability & -- & Model's architecture is readily comprehensible to users & \cite{colmenarejo2025should} \\
\addlinespace

Justifiability & Causality & XAI should justify why the system makes such a decision & \cite{lipton2018mythos} \\
\addlinespace

Legal Compliance & Regulatory compliance & Assess and increase the legal compliance of a system & \cite{langer2021we} \\
\addlinespace

Monitoring & Monitory & An XAI method should allow humans' continuing observation of model performance, also including monitoring distribution shifts and monitoring changes in the explanation (consistency) & \cite{langer2021we,navarro2021desiderata} \\
\addlinespace

Morality/Ethics & -- & Assess and increase a system's compliance with moral and ethical standards & \cite{langer2021we} \\
\addlinespace

Normativity & -- & Provide information about the system and specific information used for decision-making to justify it is embedded in a context regulated by various fields of law & \cite{langer2021we} \\
\addlinespace

Performance & Better decision making & Assess and increase the performance of a system & \cite{langer2021we} \\
\addlinespace

Privacy & -- & Assess and increase a system's privacy practices & \cite{langer2021we} \\
\addlinespace

Purposefulness & -- & The explainability requirements should be determined by several factors concerning the decision-making process, such as the degree of automation, who is the decision-maker, who is the individual affected by such a decision, in which context the decision is taken, and what are the potential effects and risks for the individuals and the society & \cite{nemec2025xai} \\
\addlinespace

Responsibility & -- & Provide appropriate means to let humans remain responsible or to increase perceived responsibility & \cite{langer2021we} \\
\addlinespace

Robustness (AI) & Consistency, Continuity & Assess and increase a system's robustness (e.g., against adversarial manipulation) & \cite{langer2021we,nemec2025xai} \\
\addlinespace

Robustness (XAI) & Consistency, Continuity & An explanation should be resilient to small input perturbations and generalize well & \cite{nemec2025xai} \\
\addlinespace

Safety & -- & Assess and increase a system's safety & \cite{langer2021we} \\
\addlinespace

Satisfaction & User experience quality & Have satisfying systems; users' overall subjective evaluation of the interaction process, its outcomes, and its associated costs (time, effort, and risk) & \cite{langer2021we,speith2023xai} \\
\addlinespace

Scalability & -- & An explanation should be practically computable for larger models & \cite{nemec2025xai} \\
\addlinespace

Science & Scientific understanding & Gain scientific insights from the system & \cite{langer2021we} \\
\addlinespace

Security & -- & Assess and increase a system's security & \cite{langer2021we} \\
\addlinespace

Sufficiency & Completeness & An explanation provides enough information to make it actionable & \cite{bender2025towards,aryal2023even} \\
\addlinespace

Transferability & -- & Make a system's learned model transferable to other contexts & \cite{langer2021we} \\
\addlinespace

Transparency & Openness & Have transparent systems & \cite{langer2021we,deters2024explainable} \\
\addlinespace

Trust & Reliance intention & Calibrate appropriate trust in the system & \cite{langer2021we,navarro2021desiderata,lipton2018mythos,rong2023towards} \\
\addlinespace

Trustworthiness & -- & Assess and increase the system's trustworthiness; ability, integrity, and benevolence & \cite{langer2021we,schlicker2025trustworthy,deters2024explainable} \\
\addlinespace

Truthfulness & -- & The information (explanation) should be accurate, truthful, and complete & \cite{langer2021we,colmenarejo2025should} \\
\addlinespace

Understandability & Comprehensibility, Intelligibility, Understanding, Intuitiveness & An explanation should be understandable to lay users & \cite{colmenarejo2025should,navarro2021desiderata,bender2025towards,nemec2025xai,hunsicker2025100216,rong2023towards} \\
\addlinespace

Usability & Ease-of-use & Have usable systems & \cite{langer2021we,speith2023xai,rong2023towards} \\
\addlinespace

Usefulness & Task utility & Have useful systems & \cite{langer2021we,speith2023xai} \\
\addlinespace

Verification & -- & Be able to evaluate whether the system does what it is supposed to do & \cite{chitgopkar2024accuracy} \\
\addlinespace

Verifiability & Coherence & An explanation should be comparable to prior expert knowledge (human's mental model) & \cite{nemec2025xai} \\ \bottomrule
\addlinespace

\caption{Vocabulary of XAI Desiderata. This table lists key desiderata we identified in the literature, providing at least one reference for each. While many more exist, and some appear under different names, we focus here on the most important and commonly cited ones.}
\label{tab:vocabulary}
\end{longtable}

\end{document}